\documentclass[10pt]{article}
\usepackage[usenames,dvipsnames]{color}

\usepackage{amsmath,amssymb}
\usepackage{mathrsfs}
\usepackage{color,graphics}
\usepackage{graphicx}
\usepackage{amsthm}
\usepackage{natbib}

\usepackage{textcomp}

\usepackage{amsbsy}
\usepackage{amsfonts}
\usepackage{latexsym}

\numberwithin{equation}{section}
\theoremstyle{plain}
\newtheorem{thm}{Theorem}[section]

\newtheorem{rmk}{Remark}[section]

\newtheorem{dfn}{Definition}[section]

\newcommand{\vir}[1]{``#1''}

\newcommand{\pg}[1]{\left\{#1\right\}}

\newcommand{\pt}[1]{\left(#1\right)}

\newcommand{\Xn}{\ensuremath{\mathcal{X}^{(n)}}}
\newcommand{\Data}{X^{(n)}}

\newcommand{\pa}[1]{\langle#1\rangle}

\newcommand{\di}{\mathrm{d}}

\def\1{\mathbf{1}}

\usepackage[T1]{fontenc}

\begin{document}

\allowdisplaybreaks

\begin{center}
{\bf \Large Empirical Bayes conditional density estimation}

\bigskip

Catia Scricciolo$^{1}$

\bigskip

{\it $^{1}$ Department of Decision Sciences, Bocconi University, Milan, Italy, catia.scricciolo@unibocconi.it}
\end{center}

\bigskip

\begin{abstract}
The problem of nonparametric estimation of the conditional density of a response,
given a vector 
of explanatory variables, is classical and of prominent importance in many 
prediction problems
since the conditional density provides a more comprehensive description of the association
between the response and the predictor than, 
for instance, does the 
regression function. The problem has applications across different fields like economy, actuarial sciences and medicine.
We investigate empirical Bayes estimation of conditional densities
establishing that an automatic data-driven selection of the prior 
hyper-parameters in infinite mixtures of Gaussian kernels, with predictor-dependent mixing weights,
can lead to estimators whose performance is on par with that of frequentist estimators
in being minimax-optimal (up to logarithmic factors) rate adaptive over
classes of
locally H\"{o}lder smooth conditional densities
and in performing an adaptive dimension reduction 
if the response is independent of (some of) the explanatory variables 
which, containing no information about the response, are irrelevant to the purpose of estimating its conditional density.
\end{abstract}

\bigskip

\section{Introduction}\label{sec:intro}

\allowdisplaybreaks

The problem of estimating the conditional density of a response, given a set of predictors,
is classical and of primary importance in real data analysis, since the conditional density
provides a more comprehensive description of the association
between the response and the predictors than, for instance, does
the conditional expectation or regression function which can only capture
partial aspects of it. The conditional density contains information on how
the different features of the response distribution, like skewness, shape and so on,
change with the covariates. Conditional density estimation for predictive purposes
have applications across different fields like economy, actuarial sciences and medicine.

Nonparametric estimation of a collection of conditional densities over a covariate space
presents two main features: 
(a) the multivariate curve may have different regularity levels along
different directions, (b)
the function may
depend only on a subset of the covariates. The goal is estimating a
multivariate function of the relevant predictors, while discarding the remaining ones,
and obtain procedures that simultaneously adopt to the unknown dimension
of the predictor and to the possibly anisotropic regularity of the function.
Classical references on nonparametric conditional density estimation taking a frequentist approach 
are
\citet{efromovich2007, efromovich2010} and \citet{hall2004}; see also the recent contribution by
\citet{riv2013}. The problem of conditional density estimation has been studied taking
a Bayesian nonparametric approach only recently and popular methods are based on generalized stick-breaking
process mixture models for which supporting results, in terms of frequentist
asymptotic properties of posterior distributions, have been given
by \citet{Pati2013456} and \citet{norets2014}. The former article provides
sufficient conditions for posterior consistency in conditional density estimation for a broad class
of predictor-dependent mixtures of Gaussian kernels.
The latter presents results on posterior contraction rates
for conditional density estimation over classes of locally (isotropic) H\"{o}lder
smooth densities using \emph{finite} mixtures of Gaussian kernels, with covariate-dependent mixing weights having a special structure. The entailed density
estimation procedure converges at a rate that
automatically adapts to the unknown dimension of the set of
relevant covariates, 
thus ultimately performing a dimension reduction,
and to the regularity level of the sampling conditional density.

In this note, the focus is on defining procedures for conditional density estimation that
attain minimax rates (up to log-factors) of posterior concentration adopting to both the dimension
of the set of relevant covariates and to the regularity level of the function. We consider a procedure based on 
\emph{infinite} mixtures of Gaussian kernels, with the same predictor-dependent mixing weights as in
\citet{norets2014}, and show that it can have a performance on par with that of the procedure
proposed by the above cited authors 
in terms of rate adaptation to the predictor dimension and to the (isotropic) regularity level. Under
the same set of assumptions on the data generating process and the prior law, the performance of the conditional density estimation procedure of an
empirical Bayesian, who considers an automatic data-driven selection of the prior hyper-parameters,
matches with that of an \vir{honest} Bayesian. We deal in detail with the isotropic case;
extension of the result to the anisotropic case follows along the same lines.

The organization of the article is as follows. Section \ref{subsec:notation} sets up
the notation. 
Section \ref{sec:mainresults} presents the main results on adaptive
empirical Bayes posterior concentration at minimax-optimal $L^1$-rates
(up to log-factors) for locally H\"{o}lder
smooth conditional densities, with contextual adaptive dimension reduction in the presence of irrelevant covariates.
Final remarks and comments are gathered in Section \ref{finalrmks}. The statement of a theorem
invoked in the proof of the main result is reported in the Appendix for easy reference.


\subsection{Notation}\label{subsec:notation}
Let $\mathbb{N}_0=\{0,\,1,\,\ldots\,\}$ be the set of non-negative integers and
$\mathbb{R}_+$ that of positive real numbers.
For any $a,\,b\in\mathbb{R}$, 
we denote by $a\wedge b$
their minimum and by $a\vee b$ their maximum.
We write \vir{$\lesssim$} and
\vir{$\gtrsim$} for inequalities valid up to a constant multiple which is universal or
inessential for our purposes. For a generic sequence
$\{a_n\}$, we use the notation $a_n=o(1)$ ($n\rightarrow\infty$) to mean that $a_n\rightarrow0$ as $n\rightarrow\infty$.
For sequences $\{a_n\}$ and $\{b_n\}$, by writing $a_n=O(b_n)$ ($n\rightarrow\infty$) we mean that
$b_n\neq0$ and there exists a constant $K>0$ so that $|a_n/b_n|<K$ for every $n\in\mathbb{N}$.

\smallskip

For $d_x\in\mathbb{N}$, let $\mathcal{X}\subseteq\mathbb{R}^{d_x}$ be the covariate space;
for $d_y\in\mathbb{N}$, let $\mathcal{Y}\subseteq\mathbb{R}^{d_y}$ be the response space and, for $d:=d_x+d_y$, let
$\mathcal{Z}=\mathcal{X}\times\mathcal{Y}\subseteq \mathbb{R}^d$ be the sample space.

For any $k\in\mathbb{N}$, if $E\subset\mathbb{R}^k$ and $x\in\mathbb{R}^k$, the translate of $E$ is the set
$E+x:=\{z+x:\,z\in E\}$. If $\xi,\,\vartheta\in \mathbb{R}^k$, 
the Euclidean distance
between $\xi$ and $\vartheta$ is $\|\xi-\vartheta\|:=\{\sum_{j=1}^k(\xi_j-\vartheta_j)^2\}^{1/2}$.

\smallskip
Let $$\mathcal{F}:=\pg{f: 
\mathcal Z\rightarrow[0,\,\infty)\,\bigg|\,
\textrm{Borel-measurable and, }\, \forall\,x\in \mathcal{X},\,\,
    \int_{\mathcal{Y}}f(y|x)\di y=1}$$
be the space of conditional probability densities with respect to Lebesgue measure $m$ on $\mathcal Y$.
The same symbol $m$ will also be used to denote Lebesgue measure on
$\mathcal Z$. 
A centered multivariate normal density with covariance matrix $\sigma^2 I$, for $I$ the
identity matrix whose dimension is clear from the context, is denoted by $\phi_\sigma$.
The symbol $\delta_z$ stands for point mass at $z$.

\smallskip

Let $Q$ be a fixed probability measure on the measurable space
$(\mathcal{X},\,\mathcal{B}(\mathcal X))$, with $\mathcal{B}(\mathcal X)$ the Borel
$\sigma$-field on $\mathcal X$, that possesses Lebesgue density $q$.

\smallskip

Given any real number $p\geq1$ and Borel-measurable function $g:\mathcal{Z}\rightarrow \mathbb{R}$,
for every $x\in\mathcal X$, we introduce the notation
$\|g\|_{p,x}:=({\int_{\mathcal{Y}}|g(x,\,y)|^p\di y})^{1/p}$ that is useful to define global distances
between conditional densities.
For any pair of (conditional) densities $f_1,\,f_2\in\mathcal F$,
let the $q$-integrated $L^1$-distance be defined as $\|f_2-f_1\|_1:=
\int_{\mathcal{X}}\|f_2-f_1\|_{1,x}q(x)\di x$ and, analogously,
the $q$-integrated Hellinger distance as
$h(f_2,\,f_1):=
\int_{\mathcal{X}}\|f^{1/2}_2-f^{1/2}_1\|_{2,x}^2q(x)\di x$.
For (conditional) densities $f,\,f_0\in\mathcal F$,
the $q$-integrated Kullback-Leibler divergence of $f$ from $f_0$ is defined as
$\textrm{KL}(f_0;\,f):=\int_{\mathcal X\times \mathcal Y} f_0q\log(f_0q/fq)\,\di m$,
$m$ being here the Lebesgue measure on $\mathcal Z$,
which coincides with the
Kullback-Leibler divergence of $fq$ from $f_0q$. Analogously, the $q$-integrated second moment of
$\log (f_0q/fq)$ is defined as
$\textrm{V}_2(f_0;\,f):=\int_{\mathcal X\times \mathcal Y} f_0q|\log (f_0q/fq)|^2\,\di m$
and coincides with the second moment of $\log (f_0q/fq)$ with respect to $f_0q$.

\smallskip

The $\epsilon$-covering number of a semi-metric space $(M,\,d)$, denoted by $N(\epsilon,\,M,\,d)$,
is the minimal number of $d$-balls of radius $\epsilon$ needed to cover the set $M$.




\section{Main Results}\label{sec:mainresults}
Let $Z^{(n)}=(Z_1,\,\ldots,\,Z_n)$ be a random sample 
of independent and identically distributed (i.i.d.)
observations $Z_i=(X_i,\,Y_i)\in \mathcal Z$, $i=1,\,\ldots,\,n$, 
from a probability measure $P_0$ on the measurable space
$(\mathcal Z,\,\mathcal B(\mathcal Z))$, where $\mathcal{B}(\mathcal Z)$ is the Borel
$\sigma$-field on $\mathcal Z$, that possesses Lebesgue density
$f_0q$ that is referred to as the \emph{true joint data generating density},
with $f_0\in\mathcal F$ the conditional density of the response $Y$, given the predictor $X$, and $q$
the marginal density of $X$, called the \emph{design density}, which
is fixed and, for theoretical investigation,
does not need to be known or estimated.
The problem is to estimate the conditional 
density 
$f_0$ 
when no parametric assumption is formulated on it,
taking an empirical Bayes approach that employs an automatic data-driven
selection of the prior hyper-parameters.
For a recent overview of empirical Bayes methods, the reader may refer to \citet{prrs2014}.
Even if the proposed empirical Bayes procedure simultaneously leads to
adaptation with respect to both aspects (a) and (b) illustrated in the Introduction,
the two issues are treated separately for ease of exposition: we first deal with adaptive estimation over classes
of locally H\"{o}lder smooth conditional densities when the dimension of the predictor is correctly specified
and then prove adaptive dimension reduction in the case where fewer covariates are relevant.
Adaptive dimension reduction clearly plays a key role in view of the curse of dimensionality.
In Section \ref{sec:dimred}, it is shown that, when
the response is independent of (some of) the covariates introduced in the model, the empirical Bayes
posterior asymptotically performs a dimension reduction,  
thus
contracting at a rate that results from the combination of the dimension of the
subset of relevant explanatory variables and the possibly anisotropic regularity level of the curve
as a function of the selected covariates.



\subsection{Empirical Bayes posterior concentration for conditional density estimation}\label{sec:rate}
In this section, we consider empirical Bayes posterior contraction rates for estimating conditional densities when the
dimension of the predictor is correctly specified. \\[-0.3cm]



\noindent\emph{Prior law specification}. A prior distribution 
can be induced on the space $\mathcal F$ of 
conditional densities by a law $\Pi_{\mathcal{X}}$
on a collection of mixing probability
measures
$\mathcal M_{\mathcal{X}}=\{P_x\in\mathcal M(\Theta),\,x\in\mathcal X\}$, where $\mathcal M(\Theta)$
denotes the space of all probability measures on some subset
$\Theta\subseteq\mathcal Y$,
using a mixture of $d_y$-dimensional Gaussian kernels to model the conditional 
density 
\[f(\cdot|x)=(F_x\ast\phi_\sigma)(\cdot)=\int_{\Theta} \phi_\sigma(\cdot-\theta)\di F_x(\theta),\quad x\in \mathcal{X},\]
where, for every $x\in\mathcal X$, $F_x$ is the cumulative distribution function
corresponding to a 
probability measure $P_x$
which is assumed to be (almost surely) discrete 
\begin{equation*}\label{eq:mixdistrib}
P_x=\sum_{j=1}^\infty p_j(x)\delta_{\theta_j(x)},
\end{equation*}
with random weights
$p_j(x)\geq0$, $j\in\mathbb{N}$, such that $\sum_{j=1}^\infty p_j(x)=1$ almost surely,
and random support points
$\{\theta_j(x)\}$ 
that are i.i.d. replicates 
drawn from a probability measure $G_x$ on $\Theta$.
Following \citet{Pati2013456}, we single out two relevant special cases. 
\begin{itemize}
\item[$\bullet$] \emph{Predictor-dependent mixtures of Gaussian linear regressions} ($\mathrm{MGLR}_x$): the conditional density is modeled as a mixture of Gaussian linear regressions
\[f(\cdot|x)=\int_{\mathbb{R}^{d_x}} \phi_\sigma(\cdot-\beta' x)\di F_x(\beta),\quad x\in \mathcal{X},\]
where 
$\beta' x$ denotes the usual inner product in $\mathbb{R}^{d_x}$ and the mixing measure $P_x$ corresponding to $F_x$
is such that $P_x=\sum_{j=1}^\infty p_j(x)\delta_{\beta_j}$ almost surely, with the vectors of regression coefficients $\beta_j\overset{\textrm{iid}}\sim G$.
For a particular structure of the random weights $p_j(x)$'s, probit stick-breaking mixtures of Gaussian kernels
are obtained. Probit transformation of Gaussian processes for constructing the stick-breaking weights
has been considered in \citet{rodríguez2011}, who exhibit applications to real data of the probit stick-breaking process model.
\\[-0.5cm]
\item[$\bullet$] \emph{Gaussian mixtures of fixed-p dependent processes}: if 
$p_j(x)\equiv p_j$ for all $x\in\mathcal X$,
we obtain
mixtures of Gaussian kernels with fixed weights. Versions of fixed-$p$ dependent Dirichlet
process mixtures of Gaussian densities (fixed $p$-DDP) have been applied to ANOVA, survival analysis and spatial modeling.
\end{itemize}

\smallskip

\noindent We consider a variant of the prior proposed in \citet{norets2014}. 
Let $\nu$ be a probability measure on $\mathcal X$ and $G$ a probability measure on
$\mathcal Y$. 
For $(\lambda,\,\tau)\in\mathcal Y\times\mathbb{R}_+$, 
with abuse of notation, let $G_\tau(\cdot-\lambda)$ denote the probability measure on $\mathcal Y$ 
with Lebesgue density $\tau^{-1}(\di G/\di m)((\cdot-\lambda)/\tau)$.
Given $(\mu_j^x,\,\mu_j^y)\in \mathcal Z$, $j\in\mathbb{N}$, and $\sigma\in\mathbb{R}_+$,
for every $x\in\mathcal X$, let
\begin{equation}\label{eq:density model}
p_{j,\sigma}(x):=\dfrac{p_j\phi_\sigma(x-\mu_j^x)}{\sum_{q=1}^\infty p_q\phi_\sigma(x-\mu_q^x)}, \quad j\in\mathbb{N}.
\end{equation}
We propose the following 
prior specification:
\[
\begin{split}
Y_i|(X_i=x_i),\, (F_x)_{x\in\mathcal X},\,\sigma&\,\sim\, (F_{x_i}\ast \phi_\sigma)(\cdot)=\sum_{j=1}^\infty p_{j,\sigma}(x_i)\phi_\sigma(\cdot-\mu_j^y),\\[-0.2cm]
\sum_{j=1}^\infty p_j\delta_{(\mu_j^x,\,\mu_j^y)}&\,\sim\,\textrm{DP}(c_0\nu \times G_\tau(\cdot-\lambda))\,\,\, \mbox{independent of}\,\,\,
\sigma\,\sim\,\textrm{IG}(\alpha,\,\beta),
\end{split}
\]
where $c_0\in\mathbb{R}_+$ is a finite constant and $\alpha,\,\beta\in\mathbb{R}_+$ are the shape and scale parameters of
an inverse-gamma prior distribution, respectively. In this case, $F_x$ corresponds to the probability measure
$P_x=\sum_{j=1}^\infty p_{j,\sigma}(x)\delta_{\mu_j^y}$.
For later use, note that, defined the mapping $g:\,x\mapsto \sum_{q=1}^\infty p_q\phi_\sigma(x-\mu_q^x)$
and modeled the conditional density $f$ as
$\sum_{j=1}^\infty p_{j,\sigma}(x)\phi_\sigma(\cdot-\mu_j^y)$, the density product $fg$
is a mixture of $d$-dimensional Gaussian densities
\begin{equation}\label{eq:passage}
f(y|x)g(x)=\sum_{j=1}^\infty p_j\phi_\sigma(x-\mu_j^x)\phi_\sigma(y-\mu_j^y).
\end{equation}
By the stick-breaking representation of a Dirichlet process (DP), the random weights $p_j=V_j\prod_{h=1}^{j-1}(1-V_h)$, $j\in\mathbb{N}$,
with $V_j\overset{\textrm{iid}}\sim\textrm{Beta}(1,\,c_0)$, and the locations
$\mu_j^y\overset{\textrm{iid}}\sim G_\tau(\cdot-\lambda)$. The last assertion is equivalent to
$\mu_j^y=\lambda+\zeta_j$, with $\zeta_j \overset{\textrm{iid}}\sim \tau^{-1}(\di G/\di m)(\cdot/\tau)$, $j\in\mathbb{N}$.
The overall prior can be rewritten as
\begin{equation}\label{eq:prior}
\begin{split}
Y_i|(X_i=x_i),\, (F_x)_{x\in\mathcal X},\,\sigma&\,\sim\, \sum_{j=1}^\infty p_{j,\sigma}(x_i)\phi_\sigma(\cdot-\lambda-\zeta_j)\\[-0.2cm]
\sum_{j=1}^\infty p_j\delta_{(\mu_j^x,\,\zeta_j)}&\,\sim\,\textrm{DP}(c_0\nu \times G_\tau)\,\,\, \mbox{independent of}\,\,\,
\sigma\,\sim\,\textrm{IG}(\alpha,\,\beta).
\end{split}
\end{equation}

For the vector $\gamma
=(\beta,\,\lambda,\,\tau^2)$ of prior hyper-parameters,
let $\Pi_{\gamma}$ stand for the product prior law $\textrm{DP}(c_0\nu \times G_\tau(\cdot-\lambda))\times\textrm{IG}(\alpha,\,\beta)$.
Let $\Pi_\gamma(B|Z^{(n)})$ denote the posterior probability of any Borel set $B$ of $(\mathcal F,\,d)$, where $d$ can be either the $q$-integrated Hellinger or $L^1$-distance. 
For any estimator $\hat\gamma_n
=(\hat\beta_n,\,\hat\lambda_n,\,\hat\tau_n^2)$ of $\gamma$ based on $Z^{(n)}$,
the empirical Bayes posterior law $\Pi_{\hat\gamma_n}(\cdot|Z^{(n)})$ is obtained by plugging $\hat\gamma_n$ into
the posterior distribution \[\Pi_{\hat\gamma_n}(\cdot|Z^{(n)})=\Pi_{\gamma}(\cdot|Z^{(n)})|_{\gamma=\hat\gamma_n}.\]
We study empirical Bayes posterior concentration rates relative to $d$
at an
ordinary smooth conditional density $f_0$, namely, we assess the order of magnitude of the radius $M\epsilon_n$
of a shrinking ball centered at $f_0$ so that
\begin{equation}\label{eq:post}
P_0^n\Pi_{\hat\gamma_n}(f\in\mathcal F:\,d(f,\,f_0)>M\epsilon_n|Z^{(n)})\rightarrow0,
\end{equation}
where $P_0^n\varphi$ is used to abbreviate expectation $\int_{\mathcal Z^{n}} \varphi \di P_0^n$ under the
$n$-fold product measure $P_0^n$. We consider the case where the true conditional density $f_0$, regarded as a mapping from $\mathcal Z$ to $\mathbb{R}_+\cup\{0\}$, satisfies a
H\"older condition in the sense of the following definition, for which we introduce some more
notation. For any $\beta\in\mathbb{R}_+$, let $\pa{\beta}:=\max\{i\in\mathbb{N}_0:\,i<\beta\}$ be
the largest non-negative integer strictly smaller than $\beta$.
For a $d$-dimensional multi-index $k =(k_1,\,\ldots,\,k_d)\in\mathbb{N}^d_0$,
define $k.=k_1 +\,\ldots\,+k_d$ and let $D^k$ denote the mixed partial
derivative operator $\partial^{k.}/\partial z_1^{k_1}\,\ldots\,\partial z_d^{k_d}$.

\begin{dfn} For any $\beta\in\mathbb{R}_+$, $\tau\geq0$ and function $L:\,\mathcal Z\rightarrow\mathbb{R}_+\cup\{0\}
$, let the class $C^{\beta,L,\tau}(\mathcal Z)$ consist of functions $f:\,\mathcal Z\rightarrow\mathbb{R}$
that have finite mixed partial derivatives $D^k f$ of all orders $k.\leq\pa{\beta}$ and,
for every $k\in \mathbb{N}_0^d$ such that $k.=\pa{\beta}$, the mixed partial derivatives of order $k$ are locally (uniformly)
H\"older continuous with exponent $\beta-\pa{\beta}$ in $\mathcal Z$ with envelope $L$,
\begin{equation}\label{eq:holder}
|(D^kf)(z+\Delta)-(D^kf)(z)|\leq L(z) e^{\tau \|\Delta\|^2}\|\Delta\|^{\beta-\pa{\beta}},\qquad \forall\,z,\,\Delta\in\mathcal Z.
\end{equation}
\end{dfn}


This function class has been previously considered by \citet{Shen01092013}, who
constructively showed that Lebesgue probability density functions in $C^{\beta,L,\tau}(\mathbb{R}^d)$
satisfying additional regularity conditions can be approximated by convolutions
with the Gaussian kernel $\phi_\sigma$ with an $L^1$-error of the order $\sigma^{\beta}$.
The construction of the mixing density in the approximation can be viewed as a multivariate extension of the results
in \citet[][ \S\,3]{kruijer2010}, the main difference being that condition
\eqref{eq:holder} is weaker than the one employed in \citet{kruijer2010}, where it is assumed that
$\log f_0\in C^{\beta,L,0}(\mathbb{R})$.

If $\epsilon_n$ is (an upper bound on) the posterior contraction rate and
the convergence in \eqref{eq:post} is at least as fast as $\epsilon_n^2$,
then $\epsilon_n$ is (an upper bound on) the rate of convergence of the
estimator $\hat{f}_n(\cdot|x)=\int_{\mathcal F} f(\cdot|x)\Pi_{\hat \gamma_n}(\di f|Z^{(n)})$. 
Since the convergence rate of an estimator cannot be faster than the minimax rate over the considered density function
class, the posterior contraction rate cannot be faster than the minimax rate.
So, if the posterior distribution achieves the minimax rate, then also
$\{\hat{f}_n(\cdot|x)\}_{x\in\mathcal X}$ has minimax-optimal convergence rate and is adaptive.

\smallskip

In order to state the main result
on empirical Bayes posterior contraction rates at locally H\"older smooth densities,
we report below the assumptions on the \vir{true} joint data generating density $f_0q$ and the prior law $\Pi_\gamma$.

\subsubsection{Assumptions on the joint data generating density and on the prior law}\label{sec:assump}

\medskip

\subsection*{Assumptions on $f_0q$}
\begin{enumerate}
  \item[($i$)] $\mathcal X=[0,\,1]^{d_x}$;\\[-0.6cm]
  \item[($ii$)] $q$ is bounded;\\[-0.6cm]

  \item[($iii$)] $f_0\in C^{\beta,L,\tau}(\mathcal Z)$. For some $\eta\in\mathbb{R}_+$,
  $\int_{\mathcal Z}(|L|/f_0)^{2+\eta/{\beta}}f_0\di m<\infty$
  and
    $$\int_{\mathcal Z}(|D^kf_0|/f_0)^{(2\beta+\eta)/k}f_0\di m<\infty \quad\mbox{ for all $k.\leq\pa{\beta}$};$$\\[-1cm]
  \item[($iv$)] there exist constants $B_0,\,\tau\in\mathbb{R}_+$ such that, for every $x\in \mathcal X$,
  $$f_0(y|x)\lesssim \exp{(-B_0\|y\|^\tau)} \quad\mbox{ for large $\|y\|$.}$$
\end{enumerate}


\subsection*{Assumption on $\Pi_\gamma$}
\begin{enumerate}
  \item[($v$)] the base probability measure $\nu\times G$ of the Dirichlet process possesses Lebesgue density 
  and there
  exist constants $p,\,C_0\in\mathbb{R}_+$ so that $$\frac{\di G}{\di m}(y)\propto \exp{(-C_0\|y\|^p)}\quad \mbox{ for large $\|y\|$}.$$
  \end{enumerate}

\medskip

Assumption ($i$) is not restrictive since we can always reduce to it by rescaling observations on the covariates
to live in the unit interval.
Assumption ($ii$) is verified as soon as the design density is continuous on the closed unit interval,
see the comments following the statement of
Theorem \ref{th:theorem1} concerning its use in the proof. Assumption ($iii$) requires H\"older type regularity of $f_0$ in addition to integrability conditions,
which jointly with assumption ($iv$), are used to approximate $f_0\1_{\mathcal X}$ with a
finite $d$-dimensional Gaussian mixture
having a sufficiently restricted number of support points, see Theorem 3, Proposition 1 and Theorem 4 of \citet{Shen01092013}.

\medskip

We now state the main result.

\begin{thm}\label{th:theorem1}
Suppose there exists a set $K_n\subset \mathbb{R}_+ \times \mathbb{R}\times\mathbb{R}_+$
such that $P_0^n(\hat\gamma_n\in K_n^c)=o(1)$. Under assumptions $(i)$-$(v)$,
the empirical Bayes posterior distribution corresponding to the
prior in 
\eqref{eq:prior} contracts at a rate $\epsilon_n=n^{-{\beta}/(2{\beta}+d)}(\log n)^t$ 
for a suitable constant $t>0$.
\end{thm}

We give a few comments on Theorem \ref{th:theorem1} before presenting its proof.
The empirical Bayes posterior distribution corresponding to the prior described in \eqref{eq:prior}
contracts at a rate $n^{-{\beta}/(2{\beta}+d)}(\log n)^t$ which differs
from the minimax $L^1$-rate attached to 
the class of locally H\"older densities $C^{\beta,L,\tau}(\mathcal Z)$ for at most a logarithmic factor.
The quality of the estimation improves
with increasing regularity level $\beta$
and deteriorates with increasing dimension $d$.
Furthermore, the rate automatically adapts to the unknown regularity level $\beta$
of the \vir{true} conditional density $f_0$, whatever $\beta\in\mathbb{R}_+$,
see, \emph{e.g.}, \citet{scricciolo2014} for an overview of the
main schemes for Bayesian adaptation. This implies existence of empirical Bayes procedures
for conditional density estimation that attain minimax-optimal
rates, up to logarithmic terms, over the full scale of locally H\"older densities
and perform as well as adaptive Bayesian procedures like the one entailed by the hierarchical
prior of finite Dirichlet mixtures of Gaussian densities proposed by
\citet{norets2014}.

The problem presents two main difficulties:
\begin{description}
\item[$(a)$] data-dependence of the prior law due to an automatic data-driven selection
of the prior hyper-parameters;\\[-0.5cm]
\item[$(b)$] dependence of 
$f_0$ on the covariates, which gives account for dependence
of the convergence rate on the dimension $d$ of the sample space $\mathcal Z$.
\end{description}
Concerning $(a)$, data-dependence of the prior can be dealt with resorting to the same key idea
as in \citet{petrone01062014} and \citet{donneteb2014}, which is based on a
prior measure change aimed at transferring data-dependence from the prior law to the likelihood, as long as
a parameter transformation can be identified.

\vspace{0.1cm}

\noindent Concerning $(b)$, dependence of $f_0$ on the covariates can be dealt with
regarding $f_0$ as a $d$-multivariate \emph{joint} density with respect to \emph{Lebesgue} measure
on $[0,\,1]^{d_x}\times \mathcal Y$. Indeed, 
$f_0$ is a \emph{joint} density, but with respect to the measure
$Q\times m$ on $\mathcal Z$, which prevents immediate use of 
Gaussian mixtures for its approximation. A device due to \citet{norets2014} based on the inequality
$$
h(f,\,f_0)\lesssim \|(fg)^{1/2}-(f_0\1_{[0,\,1]^{d_x}})^{1/2}\|_2,$$ which relates
the $q$-integrated Hellinger distance between the conditional densities $f$ and $f_0$ to the
Hellinger distance between the joint densities $fg$ and $f_0\1_{[0,\,1]^{d_x}}$,
where
$f(y|x)g(x)=\sum_{j=1}^\infty p_j\phi_\sigma(x-\mu_j^x)\phi_\sigma(y-\mu_j^y)$ by virtue of equality \eqref{eq:passage},
takes advantage of the special structure of the mixing weights
$p_{j,\sigma}(x)$ in model \eqref{eq:density model} for the conditional density $f$
to approximate the joint Lebesgue density $f_01_{[0,\,1]^{d_x}}$
by mixtures of $d$-dimensional Gaussian densities. 
Thus, the problem of approximating the \vir{true} joint data generating density $f_0q$ with $fq$ is translated into the problem of approximating
$f_01_{[0,\,1]^{d_x}}$ with mixtures of $d$-dimensional Gaussian densities.


\smallskip

\begin{proof}
We appeal to Theorem \ref{th:donnet} reported in the Appendix
which is an adapted version of Theorem 1 in \citet{donneteb2014}.

We first define the parameter transformation for the change of prior law.
For sequences $\underline{b}_n\downarrow0$, $\bar b_n\uparrow\infty$,
$\underline{l}_n\downarrow-\infty$, $\bar{l}_n\uparrow\infty$, $\underline{t}_n\downarrow0$
and $\bar{t}_n\uparrow\infty$,
consider a set
$K_n=[\underline{b}_n,\,\bar b_n)\times [\underline{l}_n,\,\bar l_n)\times [\underline{t}_n^2,\,\bar t_n^2)
\subseteq \mathbb{R}_+\times\mathbb{R}\times\mathbb{R}_+$ such that
$P_0^n(\hat\gamma_n \in K_n^c)=o(1)$.
For a sequence $u_n\downarrow0$ to be suitably defined later on, consider a $u_n$-covering
of $K_n$ by Euclidean open balls of radius $u_n$. To the aim,
let $v_n,\,w_n,\,z_n$ be positive infinitesimal sequences to be chosen as later on prescribed. Consider
\begin{description}
  \item[ -]a 
covering of $[\underline{b}_n,\,\bar b_n)$ with intervals $B_r=[b_r, \, b_{r+1})$,
where $b_r:=\underline{b}_n(1+z_n)^{r-1}$ for $r=1,\,\ldots,\,\lceil\log
(\bar b_n/\underline{b}_n)/\log(1+z_n)\rceil$,\\[-0.5cm]
  \item[ -]a $v_n$-covering of $[\underline{l}_n,\,\bar l_n)$ with intervals $L_k=[l_k,\,l_{k+1})$, where $l_k:=\underline{l}_n+(k-1)v_n$ for $k=1,\,\ldots,\,\lceil(\bar l_n-\underline{l}_n)/v_n+1\rceil$,\\[-0.5cm]
  \item[ -]a 
covering of $[\underline{t}_n^2,\,\bar t_n^2]$ with intervals $T_s=[t_s^2,\,t_{s+1}^2)$,
where $t_s^2:=\underline{t}_n^2(1+w_n)^{s-1}$ for $s=1,\,\ldots,\,\lceil2\log
(\bar t_n/\underline t_n)/\log (1+w_n)\rceil$.\\[-0.5cm]
\end{description}
For any $b\in B_r$, let $\pi_r:=b/b_r$. We have $1\leq \pi_r<1+z_n$.
For any $t^2\in T_s$, let $\rho_s:=(t^2/t_s^2)^{1/2}$. We have $1\leq \rho_s<(1+w_n)^{1/2}$.
Fix $\gamma'=(b_r,\,l_k,\,t_s^2)$. For any $\gamma=(b,\,l,\,t^2)\in B_r\times L_k\times T_s$,
the Euclidean distance 
$\|\gamma-\gamma'\|=[(b-b_r)^2+(l-l_k)^2+(t^2-t^2_s)^2]^{1/2} 
\leq
[(1+z_n)^2z_n^2\bar b_n^2+ v_n^2+ (1+w_n)^2w_n^2\bar t_n^4]^{1/2}=:u_n$. 
In order to have $u_n=o(1)$, it suffices that
$w_n=o(\bar t_n^{-2})$ and $z_n=o(\bar b_n^{-1})$.
The $u_n$-covering number $N_n$ of $K_n$ relative to the Euclidean distance
is $$N_n=O\pt{\frac{\log (\bar b_n/\underline{b}_n)}{\log (1+z_n)}\times \frac{\bar l_n-\underline{l}_n}{v_n}\times \frac{\log (\bar t_n/\underline{t}_n)}{\log (1+w_n)}},$$
with $v_n,\,w_n,\,z_n$ that need to be chosen so that $N_n=o(e^{n\epsilon_n^2})$ 
as postulated by requirement [{\bf{A1}}].

Fix $\gamma' = (b_r,\,l_k,\,t_s^2) \in B_r\times L_k\times T_s$ and consider any $\gamma=(b,\,l,\,t^2)\in B_r\times L_k\times T_s$.
If $\sigma'\sim \textrm{IG}(\alpha,\,b_r)$ then
$\pi_r \sigma' \sim \textrm{IG}(\alpha,\,b)$.
For $z_j'=(\mu_j^x,\,\zeta_j')$, if $F'=\sum_{j=1}^\infty
p_j\delta_{z_j'}\sim \textrm{DP}(c_0 \nu\times G_{t_s})$ then $F=\sum_{j=1}^\infty
p_j\delta_{(\mu^x_j,\,l+\rho_s\zeta_j')}\sim \textrm{DP}(c_0 \nu\times G_{t}(\cdot-l))$, where $l$
denotes a $d_y$-dimensional 
vector with components all equal to $l$. Throughout, we use the same symbol $l$ to denote
either the scalar or the vector, the correct interpretation being clear from the context.

Let $\theta=(F,\,\sigma)$. For every $x\in \mathcal X$, let $f_\theta(\cdot|x)=\sum_{j=1}^\infty
p_{j,\sigma}(x)\phi_{\sigma}(\cdot-\mu_j^y)$. 
The transformation $\psi_{\gamma',\gamma}(\theta)$ gives rise to the following density
\[f_{\psi_{\gamma',\gamma}(\theta)}(\cdot|x)=
\sum_{j=1}^\infty
p_{j,\pi_r\sigma'}(x)\phi_{\pi_r\sigma'}(\cdot-l-\rho_s\zeta_j').
\]

We now identify a set $B_n$ such that
\begin{equation}\label{eq:prior mass}
\inf_{\gamma\in K_n}\Pi_\gamma(B_n)\gtrsim e^{-Cn\epsilon_n^2}
\end{equation} for some 
constant $C>0$. Preliminarily, note that, by Lemma 7.1 of \citet{norets2014}, in virtue of assumption $(ii)$,
the squared $q$-integrated Hellinger distance between $f_\theta$ 
and $f_0$ 
 can be thus bounded above:
$$h^2(f_\theta,\,f_0)\leq 4\|q\|_\infty\, \|(f_\theta g)^{1/2}-(f_0\1_{\mathcal X})^{1/2}\|_2^2,$$
where $\|q\|_\infty:=\sup_{x\in\mathcal X}q(x)$ and the Lebesgue density $g$ is such that $f_\theta(y|x)g(x)=\sum_{j=1}^\infty p_j\phi_{\mu_j,\sigma}(x,\,y)$,
that is, $g(x)=\sum_{q=1}^\infty p_q\phi_{\mu_q^x,\sigma}(x)$. This allows us to use $d$-dimensional Gaussian mixtures
$\sum_{j=1}^\infty p_j\phi_{\mu_j,\sigma}(x,\,y)$ to approximate the density $f_0(y|x)\1_{\mathcal X}(x)$ defined on $\mathcal Z$.
The set $B_n$ is the same as the one described in Theorem 3.1 of \citet{norets2014}.
Let $\sigma_n=(\epsilon_n|\log\epsilon_n |^{-1})^{1/\beta}$ and
$a_{\sigma_n}=a_0|\log\sigma_n|^{1/\tau}$, with $a_0=[(8\beta+4\eta+16)/(B_0\delta)]^{1/\tau}$
for a sufficiently small $\delta>0$.
Find $b_1>\max\{1,\,1/(2\beta)\}$ so that
$\epsilon_n^{b_1}|\log \epsilon_n|^{5/4}<\epsilon_n$. As in the proof of Theorem 3.1 in
\citet{norets2014}, which is an adaptation of that of Theorem 4 in \citet{Shen01092013},
the following facts hold. First, there exists a partition $U_1,\,\ldots,\, U_K$ of $\{z\in\mathcal Z:\,\|z\|\leq a_{\sigma_n}\}$ such that,
for $j=1,\,\ldots,\,N$, with $1\leq N<K$, the ball $U_j$ is centered at $z_j=(x_j,\,y_j)$ and has diameter $\sigma_n\epsilon_n^{2b_1}$, while, for $j=N+1,\,\ldots,\,K$, each set $U_j$ has diameter bounded above by $\sigma_n$. This can be realized with
$1\leq N<K=O(\sigma_n^{-d}|\log \epsilon_n|^{d(1+1/\tau)})$.
Further extend this to a partition $U_1,\,\ldots,\, U_M$
of $\mathbb{R}^d$, for $M=O(\epsilon_n^{-d/\beta}|\log \epsilon_n|^{ds})$, with $s=1+1/\beta+1/\tau$, such that
$1\geq \inf_{(l,\,t)\in K_n}(c_0\nu\times G_t(\cdot-l))(U_j)\gtrsim (\sigma_n\epsilon_n^{2b_1})^d$
for all $j=1,\,\ldots,\,M$,
provided that $\bar l_n=O(a_{\sigma_n})$, $\bar t_n=O(a_{\sigma_n}^p)$ and $a_{\sigma_n}=O(\underline t_n|\log\epsilon_n|^{1/p})$.
Second, by virtue of assumptions $(iii)$ and $(iv)$, there exists $\theta^*=(F^*,\,\sigma_n)$, where $F^*=\sum_{j=1}^Np_j^*\delta_{\mu_j^*}$, with
$\mu_j^*=z_j$ for $j=1,\,\dots,\,N$,
so that $f_{\theta*}(y|x)g(x)=\sum_{j=1}^Np_j^*\phi_{\mu_j^*,\sigma_n}(x,\,y)$ and
$\|(f_{\theta^*}g)^{1/2}-(f_0\1_{\mathcal X})^{1/2}\|_2
=O(\sigma_n^\beta)$. Third, $P_0(\|Z\|>a_{\sigma_n})=O(\sigma_n^{4\beta+2\eta+8})$.

Let $\mathcal M(\mathbb{R}^d)$ denote the class of all probability measures on $\mathbb{R}^d$.
Define $p_j^*=0$ for $j=N+1,\,\ldots,\,M$.
Let $ B_n= \mathcal P_n\times \mathcal S_n$ be the set with
\[\begin{split}
\mathcal P_n&=\bigg\{F\in\mathcal M(\mathbb{R}^d):\,
 \sum_{j=1}^M|F(U_j)-p_j^*|\leq 2\epsilon_n^{2db_1},
  \,\,\,\min_{j=1,\,\ldots,\,M}F(U_j)\geq\epsilon_n^{4db_1}/2\bigg\}
  \end{split}\]
 and
  $\mathcal S_n=[\sigma_n(1+\sigma_n^{2\beta})^{-1/2},\,\sigma_n]$. Note that
  $M\epsilon_n^{2db_1}\leq \epsilon_n^{2d(b_1-1/2\beta)}|\log \epsilon_n|^{ds}\leq 1$ and
  $$\inf_{(l,\,t)\in \mathbb{R}\times\mathbb{R}_+}\min_{1\leq j\leq M}(c_0\nu\times G_t(\cdot-l))(U_j)^{1/2}
  \gtrsim \epsilon_n^{2db_1}(\epsilon_n^{b_1-1/2\beta}|\log \epsilon_n|)^{-d}\gtrsim \epsilon_n^{2db_1}.$$
  For every $\theta=(F,\,\sigma)\in  B_n$, the $q$-integrated Hellinger distance
  $h(f_\theta,\,f_0)=O(\sigma_n^{\beta})$. Proceeding as in Theorem 3.1 of \citet{norets2014},
  we obtain that $\max\{\mathrm{KL}(f_0;\,f_\theta),\,\mathrm{V}_2(f_0;\,f_\theta)\}=O(n\epsilon_n^2)$.
We now evaluate the probability of the set $ B_n= \mathcal P_n\times \mathcal S_n$. By applying Lemma 10
of \citet{ghosal2007},
$$\inf _{(l,\,t)\in K_n
}\mathrm{DP}_{c_0\nu\times G_t(\cdot-l)}(\mathcal P_n)
\gtrsim \exp{(-M|\log \epsilon_n|)}
\gtrsim \exp{(-c_1\epsilon_n^{-d/\beta}|\log \epsilon_n|^{ds+1})}.$$
Also, for the probability of the set $\mathcal S_n$ under the $\mathrm{IG}(\alpha,\,b)$, which is denoted by $P_b(\mathcal S_n)$, we have
\[\begin{split}
\inf_{b\in K_n}P_b(\mathcal S_n)&=\inf_{b\in K_n}\int_{\sigma_n^{-1}}
^{\sigma_n^{-1}(1+\sigma_n^{2\beta})^{1/2}}\frac{b^\alpha}{\Gamma(\alpha)}e^{-b\sigma}\sigma^{\alpha-1}\,\di \sigma\\
&\gtrsim
\underline b_n^\alpha \exp{(-\sqrt{2}\bar b_n/\sigma_n)}\sigma_n^{-\alpha}[(1+\sigma_n^{2\beta})^{\alpha/2}-1]\gtrsim \exp{(-c_2\bar b_n/\sigma_n)}
\end{split}
\]
for a suitable constant $c_2>0$, provided that $\bar b_n=O(\log^a n)$, with $a>0$,
and $\underline b_n^{-1}=O(\sigma_n^{-1})$.
Consequently, 
\[\begin{split}
\inf_{\gamma\in K_n}\mathrm{DP}_{c_0\nu\times G_t(\cdot-l)}(\mathcal P_n)\times P_b(\mathcal S_n)\gtrsim
\exp{(-c_3\epsilon_n^{-d/\beta}|\log \epsilon_n|^{(ds+1)\vee a})}\gtrsim
\exp{(-c_3n\epsilon_n^2)},
\end{split}\]
provided that, for $\epsilon_n=n^{-\beta/(2\beta+d)}(\log n)^t$, the exponent $t\geq [(ds+1)\vee a]/(2+1/\beta)$.
To complete verification of condition [{\bf{A1}}], we show that, for some constant $c_4>0$,
$$\sup_{\gamma'\in K_n}\sup_{\theta\in B_n}P_0^n\Big(\inf_{\gamma:\,\|\gamma-\gamma'\|\leq u_n}\ell_n(\psi_{\gamma',\gamma}(\theta))<-c_4n\epsilon_n^2\Big)=o(N_n^{-1}).$$
Fix $\gamma' = (b_r,\,l_k,\,t_s^2) \in B_r\times L_k\times T_s$ and consider any $\gamma=(b,\,l,\,t^2)\in B_r\times L_k\times T_s$.
For every $\theta\in B_n$,
\[\begin{split}
\inf_{\gamma:\,\|\gamma-\gamma'\|\leq u_n}f_{\psi_{\gamma',\gamma}(\theta)}(y|x)
&\geq\inf_{\gamma:\,\|\gamma-\gamma'\|\leq u_n}\sum_{j=1}^M \1_{\|\zeta_j'\|\leq a_{\sigma_n}}
p_{j,\pi_r\sigma'}(x)\phi_{\pi_r\sigma'}(y-l-\rho_s\zeta_j')\\
&\geq T_n(y)(1+z_n)^{-2}e^{-12d_xz_n/\sigma_n^2}\\&
\qquad\qquad\qquad\quad\,
\times\sum_{j=1}^M \1_{\|\zeta_j'\|\leq a_{\sigma_n}}
p_{j,\sigma'}(x)\phi_{\sigma'}(y-l_k-\zeta_j'),
\end{split}
\]
where
\[\begin{split}
T_n(y):=\exp{\Big(-\frac{1}{(\pi_r\sigma')^2}[w_n^2a_{\sigma_n}^2+d_yv_n^2+(w_na_{\sigma_n}+v_n){d_y}^{1/2}
(a_{\sigma_n}+\|y-l_k\|)]\Big)}.\end{split}
\]
Over the set $\mathcal  Y_0^n=\{(y_1,\,\ldots,\,y_n)\in(\mathbb{R}^{d_y})^n:\,\sum_{i=1}^n\sum_{j=1}^{d_y}
(y_{ij}-\mathbb{E}_0[Y_{j}])^2\leq d_yn\tau_n^2\}$, where $\tau_n=O(\log^\kappa n)$ for $\kappa>0$,
$$T_n(y)\geq\exp{\Big(-\frac{4}{\sigma_n^2}(1+d_y^{1/2})m_n[a_{\sigma_n}+4\max\{d_y^{1/2}\bar l_n/2,\,\tau_n\}]\Big)},$$
with $m_n:=\max\{w_na_{\sigma_n},\,d_y^{1/2}v_n\}$. Set $c_n(x;\,\sigma'):=\sum_{j=1}^M\1_{\|\zeta_j'\|\leq a_{\sigma_n}}
p_{j,\sigma'}(x)$, we have
$c_n(x;\,\sigma')\geq e^{-8d_x^{1/2}\epsilon_n^2} \sum_{j=1}^M\1_{\|\zeta_j'\|\leq a_{\sigma_n}}
p_j\geq e^{-8d_x^{1/2}\epsilon_n^2}(1-2\epsilon_n^{2db_1})> e^{-8d_x^{1/2}\epsilon_n^2}\epsilon_n^2$.
Let $F'$ be the distribution obtained by re-normalizing
$\sum_{j=1}^M\1_{\|\zeta_j'\|\leq a_{\sigma_n}}
p_j\delta_{(\mu_j^x,\,l_k+\zeta_j')}$.
For $\theta'=(F',\,\sigma')$, on the event $\mathcal  Y_0^n$, for a suitable constant $C'>0$,
\[\begin{split}
&\inf_{\gamma:\,\|\gamma-\gamma'\|\leq u_n}\ell_n(f_{\psi_{\gamma',\gamma}(\theta)})\\
&\qquad\qquad\geq
\sum_{i=1}^n\log \frac{f_{\theta'}(y_i|x_i)}{f_0(y_i|x_i)}-2n\log(1+z_n)+\sum_{i=1}^n\log c_n(x_i;\,\sigma')\\&
\qquad\qquad\qquad\qquad-\frac{4n}{\sigma_n^2}[(1+d_y^{1/2})m_n(a_{\sigma_n}+4\max\{d_y^{1/2}\bar l_n/2,\,\tau_n\})+3d_xz_n]\\
&\qquad\qquad\geq
\sum_{i=1}^n\log \frac{f_{\theta'}(y_i|x_i)}{f_0(y_i|x_i)}-C'n\epsilon_n^2,
\end{split}
\]
provided that $z_n=O(\sigma_n^2\epsilon_n^2)$ and $m_n=O(\sigma_n^2\epsilon_n^2(\max\{a_{\sigma_n},\,\bar l_n,\,\tau_n\})^{-1})$.
Also, we have $1-P_0^n(\mathcal Y_0^n)=O((n\tau_n^4)^{-1})$ and need that $(n\tau_n^4)^{-1}=o(N_n^{-1})$.

We show that the requirements of condition [{\bf{A2}}] are satisfied.
We start by describing a set $\mathcal F_n$ of conditional densities
such that, for some constant $\zeta>0$,
\begin{equation}\label{eq:covering}
\log N(\zeta\epsilon_n,\,\mathcal F_n,\,h)=O(n\epsilon_n^2).
\end{equation}
We consider the same sieve $\{\mathcal F_n\}$ as in Theorem 4.1 of \citet{norets2014}.
For $H_n=\lfloor n\epsilon_n^2/(\log n)\rfloor$, $\underline p_n=e^{-nH_n}$, $\underline\sigma_n=\epsilon_n^{1/\beta}$,
$\bar\sigma_n=e^{Tn\epsilon_n^2}$ for some constant $T>0$, and $\bar\mu_n=(\log n)^{\tau_1}$ for some $\tau_1>0$,
let
\[\begin{split}
\mathcal F_n&:=\bigg\{\bigg(\sum_{j=1}^{\omega} p_{j,\sigma}(x)\phi_{\sigma}(\cdot-\mu_j^y)\bigg)_{x\in\mathcal X}:\, p_j\geq \underline p_n, \,\,\mu_j^y\in[-\bar\mu_n,\,\bar\mu_n]^{d_y},\,j=1,\,\ldots,\,\omega,\,\\[0.1cm]&\hspace*{10.5cm}\omega\leq H_n,\,\,\sigma\in[\underline\sigma_n,\,\bar\sigma_n]\bigg\}.
\end{split}\]
For every fixed $\gamma'\in K_n$, 
let $\mathcal F_n(\gamma'):=\bigcup_{\gamma:\,\|\gamma-\gamma'\|\leq u_n}\psi_{\gamma',\gamma}^{-1}(\mathcal F_n)$, where
$\psi_{\gamma',\gamma}^{-1}(\mathcal F_n)$ denotes the preimage of the set $\mathcal F_n$ under
$\psi_{\gamma',\gamma}$.
We show that condition $(a)$ is satisfied. Fix any $\gamma'=(b_r,\,l_k,\,t_s^2)\in K_n$.
Proceeding as in Theorem 4.1 of \citet{norets2014}, 
\[\begin{split}
&\hspace{-0.4cm}\sup_{\gamma:\,\|\gamma-\gamma'\|\leq u_n}\sup_{\theta\in\mathcal F_n(\gamma')}\sup_{x\in\mathcal X}\|f_\theta(\cdot|x)-f_{\psi_{\gamma',\gamma}(\theta)}(\cdot|x)\|_1\\&\qquad\qquad\qquad\qquad \qquad\,\,\,\,\lesssim
\frac{1}{\sigma'(1\wedge \pi_r)}\sum_{j=1}^{d_y}[|l-l_k|+\sigma'|1-\pi_r|]+
\frac{1}{\underline\sigma_n^2}|1-\pi_r|\\&\qquad\qquad\qquad\qquad\qquad\,\,\,\,\lesssim \frac{v_n}{\underline\sigma_n}+\frac{(1+z_n)z_n\bar b_n}
{\underline\sigma_n^2\underline b_n}\lesssim \epsilon_n
\end{split}\]
as long as $v_n=O(\underline\sigma_n\epsilon_n)$
and $z_n=O(\underline\sigma_n^2\underline b_n\epsilon_n/\bar b_n)$.

Regarding condition $(b_1)$, it follows from \eqref{eq:prior mass} that
$\sup_{\gamma\in K_n}\Pi_{\gamma}(\mathcal F_n(\gamma))/\Pi_{\gamma}(B_n)\lesssim e^{Kn\epsilon_n^2/2}$
for a suitable constant $K>0$ arising from condition $(b_3)$.

To check condition $(b_2)$,
for every $\gamma'=(b_r,\,l_k,\,t_s^2)\in K_n$ and any $\theta\in \mathcal F_n(\gamma')$, we find an upper bound on
$\sup_{\gamma:\,\|\gamma-\gamma'\|\leq u_n}f_{\psi_{\gamma',\gamma}(\theta)}(\cdot|x)$ by a function (not necessarily a density)
$\bar f(\cdot|x)$.
For some constant $c_0>0$, let $a_n=c_0(\log n)^{1/\tau}$. For $\|y\|\leq a_n/2$,
if $\|\zeta_j'\|> a_n$ and $d_y^{1/2}\bar l_n\leq a_n/4$ then
$\|y-l_k-\zeta_j'\|> \|\zeta_j'\|/4$. Setting $r_n^2:=[1-16d_y^{1/2}(v_n\vee w_n)]^{-1}$, for every $\omega\leq H_n$,
\[\begin{split}
&\hspace*{-1cm}f_{\psi_{\gamma',\gamma}(\theta)}(y|x)\1_{\|y\|\leq a_n/2}(y)\\
&\quad\,\,\leq \sum_{j=1}^{\omega} p_{j,\sigma}(x)\phi_{\sigma}(y-l_k-\zeta_j')\\
&\qquad\,\,\,\,\,\,\,\,\,\times\exp{\pt{\frac{1}{\sigma^2}\max\{v_n,\,w_n\}(d_y^{1/2}+\|\zeta_j'\|)\|y-l_k-\zeta_j'\|}}\1_{\|y\|\leq a_n/2}(y)\\
&\quad\,\,\leq \max\{e^{(3/2+d_y^{1/2})^2(v_n\vee w_n)(a_n\vee \bar l_n)a_n/\sigma^2},\,r_n\}\\
&\qquad\,\,\,\,\,\,\,\,\,
\times\sum_{j=1}^{\omega} p_{j,\sigma}(x)[\1_{\|\zeta_j'\|\leq a_n}\phi_{l_k+\zeta_j',\sigma}(y)+\1_{\|\zeta_j'\|>a_n}\phi_{l_k+\zeta_j', r_n\sigma}(y)]\1_{\|y\|\leq a_n/2}(y)\\
&\quad\,\,\leq
\max\{e^{(3/2+d_y^{1/2})^2(v_n\vee w_n)(a_n\vee \bar l_n)a_n/(\pi_r\sigma')^2},\,r_n\}\\
&\qquad\,\,\,\,\,\,\,\,\, \times e^{6d_xz_n/(\sigma')^2}\1_{\|y\|\leq a_n/2}(y)\\
&\qquad\,\,\,\,\,\,\,\,\, \times \sum_{j=1}^{\omega} p_{j,\sigma'}(x)[\1_{\|\zeta_j'\|\leq a_n}\phi_{l_k+\zeta_j',
\pi_r\sigma'}(y)+\1_{\|\zeta_j'\|>a_n}\phi_{l_k+\zeta_j',r_n\pi_r\sigma'}(y)]\\
&\quad\,\,=:\bar f(y|x),
\end{split}\]
where in the third inequality we have used the fact that
$p_{j,\sigma}(x)\leq e^{6d_xz_n/(\sigma')^2} p_{j,\sigma'}(x)$.
Note that $\pi_r\sigma'\in[\underline\sigma_n,\,\bar\sigma_n]$ and
$l_k+\zeta_j'\in[-\bar\mu_n,\,\bar\mu_n]^{d_y}$ for $j=1,\,\ldots,\,\omega$, with $\omega\leq H_n$.
Set the positions
\[\begin{split}
c'&:=\max\{e^{(3/2+d_y^{1/2})^2(v_n\vee w_n)(a_n\vee \bar l_n)a_n/(\pi_r\sigma')^2},\,r_n\} \times e^{6d_xz_n/(\sigma')^2}
\end{split}\]and
\[\begin{split}
c(x)
&:=\sum_{j=1}^{\omega} p_{j,\sigma'}(x)\bigg[\1_{\|\zeta_j'\|\leq a_n}\int_{\|y\|\leq a_n/2}\phi_{l_k+\zeta_j',\pi_r\sigma'}(y)\,\di y\\&\qquad\qquad\qquad\qquad\qquad\qquad\qquad+\1_{\|\zeta_j'\|>a_n}\int_{\|y\|\leq a_n/2}\phi_{l_k+\zeta_j',r_n\pi_r\sigma'}(y)\,\di y\bigg],
\end{split}\]
and observed that 
$c(x)\leq1$ for all $x\in\mathcal X$,
under the constraints $z_n=O((n\epsilon_n)^{-2})$ and $v_n\vee w_n=O(((a_n\vee \bar l_n)a_nn\epsilon_n^2)^{-1})$,
the normalizing constant of $\prod_{i=1}^n\bar f(y_i|x_i)$
can be thus bounded above
\[\begin{split}
\prod_{i=1}^n[c'\times c(x_i)]
&< \Big(\max\{e^{(3/2+d_y^{1/2})^2(v_n\vee w_n)(a_n\vee \bar l_n)a_n/\underline \sigma_n^2},\,r_n\} \times e^{6d_xz_n(1+z_n)^2/\underline\sigma_n^2}\Big)^n\\
&\lesssim \exp{\pt{C_3(v_n\vee w_n)(a_n\vee \bar l_n)a_n(n\epsilon_n^2)^2+48d_x n z_n(n\epsilon_n^2)^2}}\lesssim e^{C_3' n\epsilon_n^2}
\end{split}\]
for suitable constants $C_3,\,C_3'>0$. Let $\mathcal Y_1=\{y\in\mathcal Y:\,\|y\|\leq a_n/2\}$.
We are allowed to consider the restriction to $(\mathcal X\times \mathcal Y_1)^n$ since, by virtue of assumption $(iv)$,
$$P_0^n((\mathcal X\times \mathcal Y_1^c)^n )=\pt{\int_0^1\int_{\|y\|>a_n/2}f_0(y|x)q(x)\,\di x\di y}^n\lesssim e^{-B_0n(a_n/2)^\tau}\lesssim e^{-B_0n\epsilon_n^2}.$$
Recalling that, in the present setting,
$\di Q_{\theta, \gamma'}/\di m=\sup_{\gamma:\,\|\gamma-\gamma'\|\leq u_n}
f_{\psi_{\gamma',\gamma}(\theta)}(\cdot|x)q(x)$, in order to show that
condition $(b_2)$ is satisfied, we need to prove that
$$\sup_{\gamma'\in K_n}\int_{\mathcal F_n^c(\gamma')}
Q_{\theta, \gamma'}^{n}(\mathcal Z^{n})\frac{\Pi_{\gamma'}(\di \theta)}{\Pi_{\gamma'}(B_n)}
=o(N_n^{-1}e^{-C_2n\epsilon_n^2}).$$ By inequality \eqref{eq:prior mass}, 
it suffices to show that 
\begin{equation}\label{eq:uBQ}
\sup_{\gamma'\in K_n}\int_{\mathcal F_n^c(\gamma')}Q_{\theta, \gamma'}^{n}(\mathcal Z^{n})
\Pi_{\gamma'}(\di\theta)=O(e^{-En\epsilon_n^2})
\end{equation} for some constant $E>(C_2\vee c_3)$, where $c_3$ plays the role of $C$ in \eqref{eq:prior mass}. The integral in \eqref{eq:uBQ}
can be thus split up:
\[
\begin{split}
&\hspace*{-0.1cm}\sup_{\gamma'\in K_n}\int_{\mathcal F_n^c(\gamma')}Q_{\theta, \gamma'}^{n}(\mathcal Z^{n})
\Pi_{\gamma'}(\di \theta)\\&
\qquad\qquad\quad=
\sup_{\gamma'\in K_n}\bigg[\int_{F\in\mathcal M(\mathbb{R}^d)}\pt{\int_{\sigma'<\underline\sigma_n}+\int_{\sigma'>\bar\sigma_n/2}}
Q_{\theta, \gamma'}^{n}(({\mathcal X}\times\mathcal Y_1)^{n})
\Pi_{\gamma'}(\di \theta)\\&\qquad\qquad\qquad\qquad\qquad\qquad\qquad\,\,\,+
\int_{F\in\mathcal F_n^c(\gamma')}\int_{\underline\sigma_n/2}^{\bar\sigma_n}
Q_{\theta, \gamma'}^{n}(({\mathcal X}\times\mathcal Y_1)^{n})
\Pi_{\gamma'}(\di \theta)\bigg]\\
&\qquad\qquad\quad=:S_1+S_2+S_3.
\end{split}\]

To deal with the term $S_1$, we partition $(0,\,\underline\sigma_n)=\bigcup_{j=0}^\infty
[\underline\sigma_n2^{-(j+1)},\,\underline\sigma_n2^{-j})$. For every $j\in\mathbb{N}_0$,
let $u_{n,j}=e_n(\underline\sigma_n2^{-j})$, with $e_n=o(1)$ so that $u_{n,j}<u_n$.
For every $\gamma'=(b_r,\,l_k,\,t_s^2)\in K_n$, consider a $u_{n,j}$-covering of $\{\gamma:\,\|\gamma-\gamma'\|\leq u_n\}$
with centering points $\gamma_i$, for $i=1,\,\ldots,\, N_j$, with $N_j\leq (u_n/u_{n,j})^3$.
For a suitable constant $A>0$,
\[
\begin{split}
&\sup_{\gamma'\in K_n}\int_{F\in\mathcal M(\mathbb{R}^d)}\int_{\sigma'<\underline\sigma_n}
Q_{\theta, \gamma'}^{n}(({\mathcal X}\times\mathcal Y_1)^n)
\Pi_{\gamma'}(\di \theta)\\&\,\,\,\,=O\bigg(\sum_{j=0}^\infty \exp{\Big(nu_{n,j}[(3/2+d_y^{1/2})^2(a_n\vee \bar l_n)a_n+6d_x]/(\underline\sigma_n2^{-(j+1)})^2+nu_{n,j}\Big)}\\[-0.1cm]
&\hspace*{5.5cm}\times
\max_{1\leq i\leq N_j} P_{b_i}([\underline\sigma_n2^{-(j+1)},\,\underline\sigma_n2^{-j})\bigg)\\[-0.1cm]
&\,\,\,\,=O\bigg(\sum_{j=0}^\infty \exp{\Big(2ne_n[(3/2+d_y^{1/2})^2(a_n\vee \bar l_n)a_n+6d_x]/
(\underline\sigma_n2^{-(j+1)})+ne_n\underline\sigma_n2^{-j}\Big)}\\[-0.1cm]
&\hspace*{5.5cm}\times
\exp{(-(\underline b_n/\underline \sigma_n)2^j)}2^{(\alpha-1)j}\sum_{i=1}^{N_j}(b_i/\underline\sigma_n)^{\alpha-1}\bigg)\\[-0.1cm]
&\,\,\,\,=O\bigg(u_n(\underline b_n/\underline \sigma_n)^{\alpha-1}\exp{(ne_n\underline\sigma_n+u_n-\log(e_n\underline\sigma_n))}\\[-0.1cm]&
\hspace*{2.9cm}\sum_{j=0}^\infty e^{-(2^j\{[\underline b_n-2ne_n[(3/2+d_y^{1/2})^2(a_n\vee \bar l_n)a_n+6d_x]/\underline\sigma_n-1\}+j(1-\alpha)\log 2)}
\bigg)\\[-0.1cm]
&\,\,\,\,=O(e^{-An\epsilon_n^2})
\end{split}
\]
provided that $e_n=o((n\underline\sigma_n)^{-1})$, $\underline b_n\gtrsim (\log n)^{-\upsilon}$ for some $\upsilon>0$ and
$e_n=O(n^{-1}(a_n\vee \bar l_n)^{-1}a_n^{-1})$.

Concerning $S_2$, for a suitable constant $B>0$ 

\[\begin{split}
S_2
&\lesssim(\max\{e^{4(3/2+d_y^{1/2})^2 (v_n\vee w_n)(a_n\vee \bar l_n)a_ne^{-2Tn\epsilon_n^2}},\,r_n\})^n\\
&\qquad\qquad\qquad\quad \times e^{24d_xnz_ne^{-2Tn\epsilon_n^2}}(1+z_n)^n\sup_{b\in K_n}P_b((\bar\sigma_n/2,\,\infty))
\lesssim e^{-Bn\epsilon_n^2}
\end{split}
\]
because 
\[\begin{split}
\sup_{b\in K_n}P_b((\bar\sigma_n/2,\,\infty))&=
\sup_{b\in K_n}\int_0^{4\bar\sigma_n^{-2}}
\frac{b_r^\alpha}{\Gamma(\alpha)}e^{-b_r\sigma}\sigma^{\alpha-1}\,\di \sigma\\
&\leq(4\bar b_n \bar\sigma_n^{-2})^{\alpha-1}(1-e^{-4\bar b_n \bar\sigma_n^{-2}})\\
&=(4\bar b_n \bar\sigma_n^{-2})^{\alpha-1}\sum_{k=1}^\infty\frac{(-1)^{k+1}}{k!}(4\bar b_n \bar\sigma_n^{-2})^k\\
&\lesssim \bar b_n e^{-2Tn\epsilon_n^2}
\exp{(-2\alpha T n\epsilon_n^2+\alpha\log \bar b_n)}
\end{split}\]
provided that $z_n=O(\epsilon_n^2)$ and $(v_n\vee w_n)=O(n^{-1}(a_n\vee \bar l_n)^{-1}a_n^{-1}\epsilon_n^2)$.

Concerning $S_3$, for any $\epsilon\in(0,\,1)$ and a suitable constant $D>0$,
\[\begin{split}
&\int_{F\in\mathcal F_n^c(\gamma')}\int_{\underline\sigma_n/2}^{\bar\sigma_n}
Q_{\theta, \gamma'}^{n}(({\mathcal X}\times\mathcal Y_1)^n)
\Pi_{\gamma'}(\di \theta)\\&\quad\,\,\,\lesssim
(\max\{e^{4(3/2+d_y^{1/2})^2 (v_n\vee w_n)(a_n\vee \bar l_n)a_n/\underline\sigma_n^2},\,r_n\})^n
e^{24d_xnz_n/\underline\sigma_n^2+nz_n}\\
&\qquad\qquad\times (1+z_n)^{-n}\underline\sigma_n^{-n}
\exp{(-ne^{-8d_y^{1/2}(v_n\vee w_n)}c\bar\mu_n^2/[2(1+z_n)^2\bar\sigma_n^2])}
\lesssim e^{-Dn\epsilon_n^2},
\end{split}\]
provided that
$z_n=O(n^{-1}\underline\sigma_n^2\epsilon_n^2)$ and $(v_n\vee w_n)=O(n^{-1}(a_n\vee \bar l_n)^{-1}a_n^{-1}\underline\sigma_n^2\epsilon_n^2)$, with
$a_n<2d_y^{1/2}\bar\mu_n$.

We now check that condition $(b_3)$ is satisfied. We show that there exists a constant $K>0$ such that,
for any fixed $\gamma'=(b_r,\,l_k,\,t_s^2)\in K_n$,
for every $\epsilon>0$ and all $\theta\in\mathcal F_n(\gamma')$ such that the $q$-integrated
Hellinger distance $h(f_\theta,\,f_0)>\epsilon$, there exists a test $\phi_n(f_\theta)$ satisfying
  \begin{equation}\label{eq:tests}
  P_0^n\phi_n(f_\theta)\leq e^{-Kn\epsilon^2} \qquad\mbox{ and }\qquad {Q_{\theta,\gamma'}^{n}}[1-\phi_n(f_\theta)]\leq e^{-Kn\epsilon^2}.
  \end{equation}
By Corollary 1 of \citet{ghosal2007}, for every $\theta\in\mathcal F_n(\gamma')$
such that $h(f_\theta,\,f_0)>M\epsilon_n$, there exists
a test $\phi_n$, which is the maximum of all tests attached to probability measures 
that are the centers of balls covering $\{\theta\in\mathcal F_n(\gamma'):\,h(f_\theta,\,f_0)>M\epsilon_n\}$, such that
\[
\begin{split}&P_0^n\phi_n\lesssim N(M\epsilon_n/4,\,\mathcal F_n(\gamma'),\,h)e^{-n(M\epsilon_n/4)^2} \qquad \mbox{ and } \qquad
\sup_{\theta\in\mathcal F_n(\gamma')}P^n_\theta(1-\phi_n)\lesssim e^{-n(M\epsilon_n/4)^2}.
\end{split}\]
By inequality \eqref{eq:covering}, the requirement on the I type error probability in \eqref{eq:tests}
is satisfied. The second requirement is satisfied provided that, for some constant $M''>0$, we have 
$h(f_{\psi_{\gamma',\gamma}(\theta)},\,f_0)>M''\epsilon_n$ for all $\gamma$ such that
$\|\gamma-\gamma'\|\leq u_n$. Since $h(f_{\psi_{\gamma',\gamma}(\theta)},\,f_0)\geq 2^{-1}(\|f_\theta-f_0\|_1-\|f_\theta-f_{\psi_{\gamma',\gamma}(\theta)}\|_1)$, it is enough
that $\sup_{x\in\mathcal X}\|f_\theta(\cdot|x)-f_{\psi_{\gamma',\gamma}(\theta)}(\cdot|x)\|_1\leq M'\epsilon_n$ for some constant $M'<M$ so that $M''= M-M'$.
This can be seen to hold as for condition $(a)$. Inequality \eqref{eq:uBQ} then follows by combining upper bounds on $S_1,\,S_2$ and $S_3$.

The proof is completed noting that the assertion follows by choosing sequences $v_n$, $w_n$ and $z_n$ so that all
the constraints arisen in the proof are simultaneously satisfied.
\end{proof}

\medskip

\begin{rmk}
Theorem \ref{th:theorem1} takes into account only a data-driven choice of the scale parameter of an inverse-gamma prior
on the bandwidth, but an empirical Bayes selection of the shape parameter
could be considered as well. In order to identify the mapping for
the change of prior measure, it suffices to note that, for $\alpha\in\mathbb{N}$,
if $\alpha_r\overset{\mathrm{iid}}\sim \mathrm{Gamma}(1,\,1)$, $r=1,\,\ldots,\,\alpha$, then
$\beta/(\sigma_1+\,\ldots\,+\sigma_\alpha)\sim\mathrm{IG}(\alpha,\,\beta)$.
\end{rmk}



\subsection{Empirical Bayes dimension
reduction in the presence of irrelevant covariates
}\label{sec:dimred}
We now deal with the case where a $d_x$-dimensional explanatory variable is
considered, but not all the covariates
are relevant to the response whose conditional distribution may depend only on fewer of them,
say $0\leq d_x^0\leq d_x$, which, without loss of generality, can be thought of as the first $d_x^0$ of the whole collection
employed in the model specified in \eqref{eq:prior}.
Besides rate adaptation, another appealing feature of the empirical Bayes procedure herein considered
is automatic dimension reduction in the presence of irrelevant covariates,
on par with the posterior distribution corresponding to the prior proposed by \citet{norets2014}.
The posterior automatically
selects the model with the subset of relevant covariates among all competing models.

\begin{thm}\label{th:theo}
Suppose that the true conditional density $f_0$ depends on the first $d_x^0\in\mathbb{N}_0$
covariates and satisfies assumptions $(iii)$-$(iv)$ of Section \ref{sec:assump}.
Under the same conditions as in Theorem \ref{th:theorem1},
the empirical Bayes posterior distribution corresponding to the
prior in \eqref{eq:prior} contracts at a rate $\epsilon_n=n^{-{\beta}/(2{\beta}+d^0)}(\log n)^t$, with
$d^0:=d_x^0+d_y$ and $t>0$ a suitable constant.
\end{thm}

The proof follows the same trail as that of Theorem \ref{th:theorem1}, the only
difference arising from the prior concentration rate
which turns out to depend on the dimension $d_x^0$ of the relevant covariates of $f_0$
because, for all the locations of the approximating Gaussian mixture, when $k>d_x^0$,
the components $\mu_{jk}^x=0$ so that eventually the
mixture does not depend on the covariates $x_k$ for $k=d_x^0+1,\,\ldots,\,d_x$.

As a simple consequence of Theorem \ref{th:theo},
we have that, if $d_x^0=0$, then $f_0(y|x)=f_0(y)$ and the response is stochastically independent
of the predictor.


\section{Final Remarks}\label{finalrmks}

In this note, we have proposed an empirical Bayes procedure for conditional density estimation
based on infinite mixtures of Gaussian kernels with predictor-dependent mixing weights
and have shown that a data-driven selection of the prior hyper-parameters 
can lead to inferential answers that are similar,
for large sample sizes, to those of hierarchical posteriors in automatically adapting to the
dimension of the set of relevant covariates and to the regularity level of the true sampling conditional density.
An empirical Bayes selection of the prior hyper-parameters
leads to pseudo-posterior distributions with the same performance as fully Bayes posteriors, provided the estimator $\hat\beta_n$
of the scale parameter of an inverse-gamma prior on the bandwidth takes values in a set $[\underline b_n,\,\bar b_n)$ such that
$P_0^n(\hat\beta_n\in [\underline b_n,\,\bar b_n)^c)=o(1)$, a requirement that imposes restrictions on the sequences $\underline b_n$ and $\bar b_n$,
in particular, on the decay rate at zero of $\underline b_n$, which is expectedly more
important than that at which $\bar b_n\uparrow\infty$. 
If
the prior hyper-parameter has an impact on posterior contraction rates,
then the choice of the plug-in estimator is crucial and requires special care.
This may, for example, rule out the maximum marginal likelihood estimator for $\beta$.
When the hyper-parameter does not affect posterior contraction rates, as it is the case for the mean $\lambda$ and
variance $\tau^2$ of the Dirichlet base measure, there is more flexibility
in the choice of the estimator: different choices are indistinguishable in terms of the posterior behavior
they induce and empirical Bayes posterior contraction rates are the same as those of any posterior corresponding to a prior
with fixed hyper-parameters.

The result of Theorem \ref{th:theo} deals with isotropic H\"older densities but
an extension to anisotropic densities, that have different levels of regularity
along different directions, is envisaged.
 In the anisotropic case, the presented results provide
adaptive rates corresponding to the least smooth direction.
Sharper rates can be obtained along the lines
of Section 5 in \citet{Shen01092013} combined with the preceding treatment
using component-specific bandwidths. Details are omitted.

\begin{centering}
\section*{APPENDIX}
\end{centering}

In this section, an adapted version of Theorem 1 in \citet{donneteb2014} is reported for easy reference.
Some additional notation is preliminarily introduced.

\medskip

Let $(\Xn,\,\mathcal B_n,\, (P_\theta^{(n)}:\,\theta \in \Theta))$ be a sequence of statistical experiments,
where $\Xn$ and $\Theta$ are Polish spaces endowed with their Borel $\sigma$-fields $\mathcal B_n$ and $\mathcal B(\Theta)$, respectively.
Let $d(\cdot,\,\cdot) $ denote a (semi-)metric on $\Theta$.
Let $\Data\in \Xn$ be the observation at the $n$th stage from $P_{\theta_0}^{(n)}$, where $\theta_0$ denotes the true parameter.
Let $\mu^{(n)} $ be a $\sigma$-finite
measure on $(\Xn,\,\mathcal B_n)$ dominating all probability measures $P_\theta^{(n)}$, for $\theta \in \Theta$.
For every $\theta\in\Theta$, let $\ell_n(\theta)$ denote the log-likelihood ratio
$\log (p_\theta^{(n)}/p_{\theta_0}^{(n)})$.

We consider a family of prior distributions $\{\Pi_\gamma\}$ on
$(\Theta,\,\mathcal B(\Theta))$, with $\Gamma \subseteq \mathbb{R}^k$, $k \in\mathbb{N}$.
Let $\Pi_\gamma( \cdot|\Data)$ stand for the posterior distribution corresponding to 
$\Pi_\gamma$.
For any measurable function $\hat\gamma_n:\,\Xn\rightarrow\Gamma$,
the empirical Bayes posterior law $\Pi_{\hat\gamma_n}(\cdot|X^{(n)})$ is obtained by plugging $\hat\gamma_n$ into
the posterior distribution, $$\Pi_{\hat\gamma_n}(\cdot|X^{(n)})=\Pi_{\gamma}(\cdot|X^{(n)})|_{\gamma=\hat\gamma_n}.$$

The statement of the theorem follows.

\bigskip

\begin{thm}[\citet{donneteb2014}]\label{th:donnet}
Let $\theta_0\in \Theta$. For every $\gamma,\,\gamma'\in\Gamma$,
let $\psi_{\gamma, \gamma'}:\,\Theta\rightarrow\Theta$ be a measurable mapping such that,
if $\theta\sim\Pi_\gamma$, then $\psi_{\gamma, \gamma'}(\theta)\sim \Pi_{\gamma'}$. Assume that 

\begin{itemize}
\item[{\bf [A1]}] there exist sets $K_n\subseteq\Gamma$ with $P_{\theta_0}^{(n)}(\hat\gamma_n\in K_n^c)=o(1)$,
positive sequences $u_n,\,\epsilon_n\downarrow0$, with $n\epsilon_n^2\rightarrow\infty$,
for which $N_n:= N(u_n,\,K_n,\,\|\cdot\|)=o(e^{n\epsilon_n^2})$ and sets $B_n\in\mathcal B(\Theta)$ such that,
for some constant $C_1>0$, 
$$
\sup_{\gamma\in K_n} \sup_{\theta \in B_n}
P_{\theta_0}^{(n)}\Big( \inf_{\gamma': \ \|\gamma' -\gamma\| \leq u_n} \ell_n(\psi_{\gamma, \gamma'}(\theta))
< -C_1n\epsilon_n^2  \Big) = o(N_n^{-1});
$$
\end{itemize}
\medskip

\noindent{\bf [A2]} for every $\gamma \in K_n$, there exists a set $\Theta_n(\gamma)\in\mathcal B(\Theta)$
such that
\begin{itemize}
\item[$(a)$] $\sup_{\gamma':\,\|\gamma'-\gamma\|\leq u_n}\sup_{\theta \in \Theta_n(\gamma) } d( \theta,\, \psi_{\gamma,\gamma'}(\theta))\leq M'\epsilon_n$
for some constant $M'>0$, \hfill
\item[$(b)$] for constants $\zeta,\,K>0$ and $C_2>C_1$,\\
$(b_1)$ $\log N(\zeta \epsilon_n,\, \Theta_n(\gamma),\,d) \leq K n\epsilon_n^2 /2\, \hfill\mbox{ and }\, \hfill\sup_{\gamma \in  K_n} \dfrac{ \Pi_\gamma(\Theta_n(\gamma)) }{ \Pi_\gamma(B_n) }  \leq e^{ K n \epsilon_n^2/2},$\\[5pt]
$(b_2)$ defined 
$Q_{\theta, \gamma}^{(n)}$ such that $\di Q_{\theta, \gamma}^{(n)}/\mathrm{d}\mu^{(n)}:= \sup_{\gamma':\,\| \gamma'  - \gamma\| \leq u_n} p_{\psi_{\gamma, \gamma'}(\theta)}^{(n)},$\\[5pt]
\[
\sup_{\gamma \in K_n} \int_{\Theta\setminus\Theta_n(\gamma)}
Q_{\theta, \gamma}^{(n)}(\Xn)\frac{\Pi_\gamma(\di\theta)}{\Pi_\gamma(B_n)}=o(N_n^{-1}e^{-C_2n\epsilon_n^2}),
\]

$(b_3)$ for any 
$\epsilon>0$, $\theta \in \Theta_n(\gamma)$ with $d(\theta,\, \theta_0)>\epsilon$, there exists a test $\phi_n(\theta)$ with\\[-4pt]

\hspace*{0.5cm} $P_{\theta_0}^{(n)}\phi_{n}(\theta)\leq e^{-K n \epsilon^2 }$ \hfill and \hfill $Q_{\theta, \gamma}^{(n)}
[1-\phi_n(\theta)] \leq e^{-K n \epsilon^2 }$.
\end{itemize}

\medskip

\noindent Then, for a sufficiently large constant $M>0$,
$$P_{\theta_0}^{(n)}\Pi_{\hat\gamma_n}\big(d(\theta,\,\theta_0)>M\epsilon_n|X^{(n)}\big)\rightarrow0.$$
\end{thm}



\section*{Acknowledgements}
The author would like to thank the Editors and two anonymous referees
for their remarks and comments which helped to
improve the initial version of the manuscript.
Bocconi University is gratefully acknowledged for providing financial support.

\bibliographystyle{statistica}
\bibliography{bibdatabase}

\end{document}